# Data-driven Inferences of Agency-level Risk and Response Communication on COVID-19 through Social Media based Interactions

**Md Ashraf Ahmed**
Ph.D. Student
Department of Civil and Environmental Engineering
College of Engineering and Computing
Florida International University
10555 West Flagler Street, EC 2900, Miami, FL 33174
Email. mpave002@fiu.edu

**Arif Mohaimin Sadri, Ph.D.**
Assistant Professor
Moss Department of Construction Management
College of Engineering and Computing
Florida International University
10555 West Flagler Street, EC 2934, Miami, FL 33174
Email. asadri@fiu.edu
(Corresponding Author)

**M. Hadi Amini Ph.D.**
Assistant Professor
School of Computing and Information Sciences
College of Engineering and Computing
Florida International University
11200 SW 8th Street, CASE 354, Miami, FL 33199
Email. moamini@fiu.edu

Word Count: 01 tables + 5 figures + 6,200 words = 6,450 word equivalents

**ABSTRACT**

Risk and response communication of public agencies through social media played a significant role in the emergence and spread of novel Coronavirus (COVID-19) and such interactions were echoed in other information outlets. This study collected time-sensitive online social media data and analyzed such communication patterns from public health (WHO, CDC), emergency (FEMA), and transportation (FDOT) agencies using data-driven methods. The scope of the work includes a detailed understanding of how agencies communicate risk information through social media during a pandemic and influence community response (i.e. timing of lockdown, timing of reopening) and disease outbreak indicators (i.e. number of confirmed cases, number of deaths). The data includes Twitter interactions from different agencies (2.15K tweets per agency on average) and crowdsourced data (i.e. Worldometer) on COVID-19 cases and deaths were observed between February 21, 2020 and June 06, 2020. Several machine learning techniques such as (i.e. topic mining and sentiment ratings over time) are applied here to identify the dynamics of emergent topics during this unprecedented time. Temporal infographics of the results captured the agency-levels variations over time in circulating information about the importance of face covering, home quarantine, social distancing and contact tracing. In addition, agencies showed differences in their discussions about community transmission, lack of personal protective equipment, testing and medical supplies, use of tobacco, vaccine, mental health issues, hospitalization, hurricane season, airports, construction work among others. Findings could support more efficient transfer of risk and response information as communities shift to new normal as well as in future pandemics.

**INTRODUCTION AND MOTIVATION**

No person or place is safe from natural or man-made hazards or losses resulting from an extreme event. Infectious disease can significantly add to the far-reaching adverse consequences and sufferings of countries and communities. One way to reduce such impacts can be done by investing in communication networks similar to the resilience observed in civil infrastructure systems (*1*). However, advancing such resilience capacities is not that simple since often times it is quite challenging to identify low cost alternatives. Communicating risks, connecting community networks (*2*), and promoting resilience culture can aid local communities in making progress to increase their resilience. As such, community coalitions of elected representatives from the public and private sectors, with ties and funding from federal and state governments, and with input from local residents, are becoming very relevant. Such coalitions may be reinforced to determine the exposure and vulnerability of the community to hazard, to inform and communicate risk, and to analyze and extend the capacity of the community to manage such risks. A truly robust coalition will have a clear leadership and governance system at the core, and people with sufficient time, expertise, and commitment required to establish and sustain relationships among all the community stakeholders (*3*).

Public agencies engagement in risk communication can lead to more effective decision-making and enhanced public feedback to the regulatory process. Risk and response communication of public agencies through social media played a significant role in the emergence and spread of novel Coronavirus (COVID-19) and such interactions were echoed in other information outlets. The primary goal of this study is to mine and analyze time-sensitive agency-level social media data (rich spatio-temporal data) as well as crowd-sourced data (i.e. Worldometer). The scope of the work includes a detailed understanding of how agencies communicate risk information through social media during a pandemic and influence community response (i.e. timing of lockdown, timing of reopening) and disease outbreak indicators (i.e. number of confirmed cases, number of deaths). The specific aims are twofold: (1) to document how public agencies interact and communicate health risk information through their online social networks during a major disease outbreak; (2) to examine how online social networks influence the current COVID-19 pandemic situation in terms of the change in daily number of cases and deaths in United States.

To achieve the goal and aims, the study have utilized social media interactions of four major public agencies e.g. World Health Organization (WHO), Center for Disease Control and Prevention (CDC), Federal Emergency Management Agency (FEMA) and Florida Department of Transportation (FDOT) with crowd-source information on COVID-19. The data includes Twitter interactions from these agencies (2.15K tweets per agency on average) and crowdsourced information about COVID-19 cases and deaths between February 21, 2020 and June 06, 2020. Several machine learning techniques such as (i.e. topic mining and sentiment ratings over time) are applied here to identify the dynamics of emergent topics during this unprecedented time. The contribution to the knowledge gaps based on empirical literature are listed below:

- It has developed methods to advance our understanding of how public agencies influence online, while communicating health risks and interacting in their respective communities as the disease continues to spread;
- It informed the literature on how information is exchanged among people who are socially connected online and exposed to health risk in such outbreaks of disease;
- This study used novel machine learning techniques to identify the dynamics of risk communication strategies during this unprecedented time ;

- It generated temporal infographics of the results captured the agency-levels variations over time in circulating information

## LITERATURE REVIEW

The transmutability of the latest coronavirus disease in 2019 was tracked in Zhejiang, China accounting for transmissions from imported cases. While Zhejiang is one of the worst 21 affected provinces, an interruption in the transmission of the disease (i.e. an instant reproduction number < 1) was observed in early / mid-February following an early social-distancing response to the outbreak (*4*). Zhang et al. developed data-driven, susceptible-exposed-infectious-quarantine-recovered (SEIQR) models to simulate the outbreak of Coronavirus with measures of social distancing and epicenter lockdown. Population migration data combined with officially recorded data were used to estimate model parameters and then to measure the daily exported infected individuals by estimating the daily infected ratio and the daily susceptible population size (*5*).

In 2003, public health interventions were crucial in the prevention of the SARS epidemic. Community containment requires steps ranging from increasing social distancing to city-wide quarantine. Whether these steps (isolation, quarantine, social distancing) will be appropriate to monitor 2019-nCoV depends on resolving some unanswered issues (*6*). Many countries only seek to achieve social distancing and hygiene measures when widespread transmission is apparent. It gives the virus a number of weeks to propagate with a higher basic reproductive number than if it had been in place before transmission was observed or widespread. Hence, preventive, low cost, enhanced hygiene and social distance in the sense of imminent population transmission of the novel coronavirus COVID-19 should be considered (*7*).

Social media is considered as an emerging platform for efficient crisis communication in recent literature (*8; 9*). Traditional media is primarily intended for one-way communication whereas social media allows two-way communication, hence social media platforms are exclusively different (*10*). Sadri et al. explained the critical role of social media during crisis by facilitating communication and information dissemination to both evacuee and non-evacuee during hurricane Sandy (*11*). Zhang et al. envisages intelligent public disaster information and alert based on social media, which has three functions: (1) effectively and efficiently collecting disaster situational awareness information; (2) promoting self-organized assistance activities; and (3) enabling emergency management agencies to hear from the public. The results of this analysis highlight the importance of such research fields; (1) a fine-grained social media catastrophe ontology with semantic interoperability, (2) trend knowledge network pattern and emerging prominent users, (3) fine-grained societal impact assessment due to infrastructure failures, and (4) best practices for social media use during disasters (*12*).

Austin et al. examines how audiences seek social and traditional media information, and what factors influence media usage during crises. Using the model of social-mediated crisis communication (SMCC), a review of crisis information and sources reveals that audiences use social media for insider information and check-in with family / friends during crises, and use traditional media for educational purposes. Convenience, interaction, and personal feedback encourage the use of conventional and social media; usage of both discourages overload. Humor and beliefs towards social media uses discourage the use of social media while legitimacy promotes the use of conventional media. The findings stressed the importance of the influence of third parties in crisis communication and the need to use traditional and social media in crisis response (*13*).

Freimuth et al. describes the design, implementation, and assessment of a risk communication simulation during the first hours of a pandemic. The simulation design was focused on the communication of crisis and emergency risks principles upheld by the Centers for Disease Control and Prevention (CDC), as well as the author's collective experience. Several local health district risk communicators in Georgia responded to a scenario where after returning from an international conference, every community in the state had teenagers infected with avian flu. The evaluation revealed that, under the time pressures of a realistic and stressful event, local risk communicators had much greater difficulty following the principles of risk communication than they did in a tabletop workout. Strengths and weaknesses of local risk communicators' performance are identified in addition to the lessons learned on designing and implementing a simulation for risk management (*14*).

Palen et al. (*15*) studied the rapid growth of social media in a number of disaster situations, exploring issues such as citizen engagement, community-oriented computing, distributed problem solving, and digital volunteerism as modes of socio-technical innovation, as well as issues of situational knowledge and truthfulness as opportunities and challenges emerging from the social media data deluge. The chapter also discusses the study that looks at integrating social media technologies and data into current emergency response work. Reflecting on the decade-old area of science, the authors warned of the danger that all "crisis" encounters can fail unintentionally without differentiation, which appears to happen because social media networks cross-cut all emergency situations. Through an effort to isolate what social media adds new, there is a tendency to fail to recognize how non-technological influences on cultural socio-behavioral scales greatly affect the usage of social media itself.

Misinformation spreading in social media are also becoming an evolving concern. Monahan et al. mentioned that mass media play an important but often misunderstood role in the events of a catastrophe. Research has repeatedly shown that disaster-related media reporting appears to be riddled with disinformation and promotes misconceptions about race, social status, aggression and crime. Studies have found that powerful media campaigns can support prevention efforts, strengthen early warning systems, facilitate orderly and prompt evacuation procedures and help bring communities together in times of upheaval. Authors review research on the relationship between media and disaster to highlight the many ways media can positively impact disaster preparation and recovery, while also highlighting the many issues associated with disaster reporting. Future directions for media-disaster research are being discussed along with ways in which media staff and emergency response professionals can handle the media-disaster relationship more efficiently before, during and after emergency incidents (*16*).

Battur et al. detected twitter bot, which is a software that sends fake tweets automatically to users. Detecting bots is necessary to identify the fake users and to protect the genuine users from misinformation and malicious intents. The study proposes an approach to detect the twitter bots using several machine learning algorithms; such as Decision Tree, Multinomial Naïve Bayes, Random Forest and Bag of Words. The algorithm with highest accuracy (Bag of Words) is used to test real time data (*17*). Asr et al. stated that misinformation detection at the level of full news articles is a text classification problem and reliably labeled data in this domain is rare. Previous work relied on news headlines, microblogs, tweets and articles collected from so-called

"reputable" and "suspicious" websites and labeled accordingly. Authors leveraged fact-checking websites to collect individually labeled news articles with regard to the veracity of their content (*18*). Huang et al. proposed a systematic meta-analysis (SMA) looked at 38 studies involving real responses to hurricane warnings and 11 studies with expected responses to hypothetical hurricane performed since 1991 (*19*).

## METHODOLOGY

The focus of this study is to examine how public agencies risk assessments, risk averting behaviors, and crisis communication patterns in online social network influence the pandemic situation. To reveal the interaction patterns, several machine learning algorithms have used in this study. At first, sentiment analysis over time for major public organizations have performed. Then Dynamic Topic Models (based on Topic Model theory) was applied to identify the topics over the same timeline which causes the change of sentiments.

Sentiment analysis is a method of Natural Language Processing (NLP) task at many levels of granularity. Starting from a document level classification task, it has been handled at the sentence level and more recently at the phrase level (*20*). It is well recognized that twitter user-generated content with rich sentiment information should be utilized for many applications such as search engines and other information systems. While tweet level sentiment analysis results indeed provide very useful information, the overall or general sentiment tendency towards topics are more appealing in some scenarios. For example, people are curious about how others feel about Apple's new product, "iPhone11," and it will offer great convenience for them if major opinions are collected from massive tweets (*21*).

A topic model is a statistical model to explore the abstract "topics" that happen in an assembly of information in machine learning and natural language processing. This was first explored by David Blei according to the most common topic model named Latent Dirichlet Allocation (LDA). The instinct behind LDA is that the set of texts reveal numerous topics. In topic models, first, the algorithm chose a topic, then sample a set of words from the given topic. Clusters of similar terms are the themes or topics created by topic modelling technology. A topic model is a recurrently performed text-mining tool for the discovery of hidden semantic structures in a text body. Topic models can help us to organize and provide insights into understanding large collections of unstructured text bodies (*22*).

The dynamic topic model (DTM) includes a group of probabilistic time series model, which is used to observe the time evolution of topics in huge document collections. This group of models was proposed by David Blei and John Lafferty and is an extension to Latent Dirichlet Allocation (LDA) that can handle chronological documents. In LDA, both the order and the words appear in a document, whereas words are still assumed to be interchangeable, but in a dynamic topic model, the order of the documents plays a key role. The method is to use state-space models to represent the topics on the natural parameters of the multinomial distributions. To perform approximate posterior inference over the latent topics, variation in approximations based on Kalman filters and nonparametric wavelet regression are developed. In addition to providing sequential, quantitative, and predictive models, DTM provides a qualitative window into the contents of a large document collection.

It is assumed that the data is divided by time slice in a dynamic topic model, for example, by month. It is modeled as each slice's documents with a $K$ component subject model, where slice

*t*-related topics evolve from slice *t-1*-related topics. Let $\beta_t$ denote the *V*-vector of natural parameters for topic *k* in slice *t* for a *K*-component model with *V* terms. A multinomial distribution is usually represented by its mean parameterization. If we denote the mean parameter of a *V*-dimensional multinomial by $\pi$, the mapping $\beta_i = log\ (\pi_i/\pi_V)$ of the *i*th element of the natural parameter is given. Dirichlet distributions are used in typical language modeling applications to model uncertainty over word distributions. The Dirichlet, however, is not conducive to sequential modeling. Alternatively, we chain the natural parameters of each $\beta_t$ subject into a state-space model that evolves with Gaussian noise; equation (2) shows the simplest version of such a model.

$$\beta_t|\beta_{t-1} \sim N(\beta_{t-1}, \sigma^2 I) \tag{1}$$

In LDA, the document-specific topic proportions $\theta$ are drawn from a Dirichlet distribution. In the dynamic topic model, a logistic normal is used with mean α to express uncertainty over proportions. The sequential structure between models is again captured with a simple dynamic model is expressed by equation (3).

$$\alpha_t|\alpha_{t-1} \sim N(\alpha_{t-1}, \sigma^2 I) \tag{2}$$

For simplicity, it did not model the dynamics of topic correlation, as it was done for static models by Blei and Lafferty (*23*). By chaining together topics and topic proportion distributions, it has sequentially tied a collection of topic models.

## DATA SOURCE AND DATA COLLECTION

Traditional datasets have limited capacity to adequately capture user risk communication strategies and analyze user concerns with such details and coverage. As such, social media datasets, enriched with user activity information, will be useful to capture user sentiments and concerns in real time and help early detection of the people exposed to health-risks in the vulnerable communities. Twitter, in particular, provides unique features to release their data through their Application Programming Interface (API) and make it publicly available which could then be combined with other complementary information (e.g. crowdsourced data) specially over the timeline of COVID-19 crisis. For this study, Twitter Search APIs is used to collect and store public agencies crisis interactions through social media outlets. Natural language processing and machine learning techniques are adopted here to extract user concerns, response, and needs over time. Besides, Worldometer data (crowdsource data) is used for extracting the daily number of cases and deaths due to COVID-19 for United States.

The goal of the study is to reveal different agencies perspective during COVID-19 that exists in social media interactions. To achieve this goal, the first key data source we considered is twitter data. We were particularly interested in the tweets generated from major public health agencies (WHO and CDC), disaster management (FEMA) and transportation agencies (FDOT). Hence, we collected historical tweets during the COVID-19 pandemic of these four major agencies to perform the analysis. The summary of the collected twitter data is given below, which shows that WHO was the most active on Twitter than any other organization. All the twitter data were collected from 02.21.2020 to 06.06.2020 which is around three and half months.

- World Health Organization (WHO) Tweets- 4,434 tweets
- Centers for Disease Control and Prevention (CDC) Tweets- 868 tweets

- Federal Emergency Management Agency (FEMA) Tweets- 1,996 tweets
- Florida Department of Transportation (FDOT) Tweets- 1,262 tweets

## ANALYSES AND RESULTS

To reveal the interaction patterns of the public organizations, several machine learning algorithms; e.g. sentiment analysis and topic frequency over time are applied here. The graphical representations and the result interpretations of the tweets from four public organizations are listed below.

### WHO Interactions

To understand the dynamics of communication pattern, 15 optimum number of topics were identified from static topic model analysis of WHO tweets. In Figure 1a, the topic frequencies are plotted along with the time and in Figure 1b, average sentiment score of tweets are plotted over the same timestamp. WHO emphasized on community transmission and the need of vaccine in March; lack of nurses was discussed (Figure 1a) at the beginning of April; shortage of PPE, need of staying home and importance of wearing masks became prominent from late April and in May; all of these should be emphasized from January (beginning of the pandemic). Requirement of Government support got notified in the beginning of May. These late responses show the clear deficiency in preparedness to handle the pandemic situation. Also, the number of cases (COVID-19 in countries topic) found increasing after 1.5 months' time interval. Several other topics such as importance of health concerns, responses in different countries, social measures, and use of tobacco got attention also from WHO in twitter.

By comparing Figure 1a and Figure 1b (rolling mean or 1-day running average), it can be said that the tweet sentiment went to most negative while WHO emphasized on alarming number of cases of COVID-19, lack of peoples' responses, shortage of PPE and negative impact of using tobacco. Then, positive sentiments have found when WHO discussed about need of government support, responses from people, development of vaccine, importance of home quarantine and use of tobacco to develop the vaccine. Hence, the use of tobacco considered as negative topic in the beginning, but later it turned into positive topic.

### CDC Interactions

From CDC tweets, 8 optimum number of topics are plotted over the three and half month timeline in Figure 2a. CDC put importance on following their guideline consistently, increase in number of cases in whole March, hospitalization rates at the end of April which also found frequent after one month, pandemic stress in May rather than from the beginning, prevention of spread during mid-April, contact tracing at the beginning of April and risk of older people in March and May.

After comparing Figure 2a and Figure 2b, positive spikes in tweet sentiment have found when CDC put emphasize on following their guideline from the beginning and mostly on May, importance of contact tracing at the beginning of April, reducing the spread of the virus as well as the pandemic stress. Besides, negative sentiments are observed when elderly people's risk, increased number of cases and hospitalization rates became more frequent in twitter.

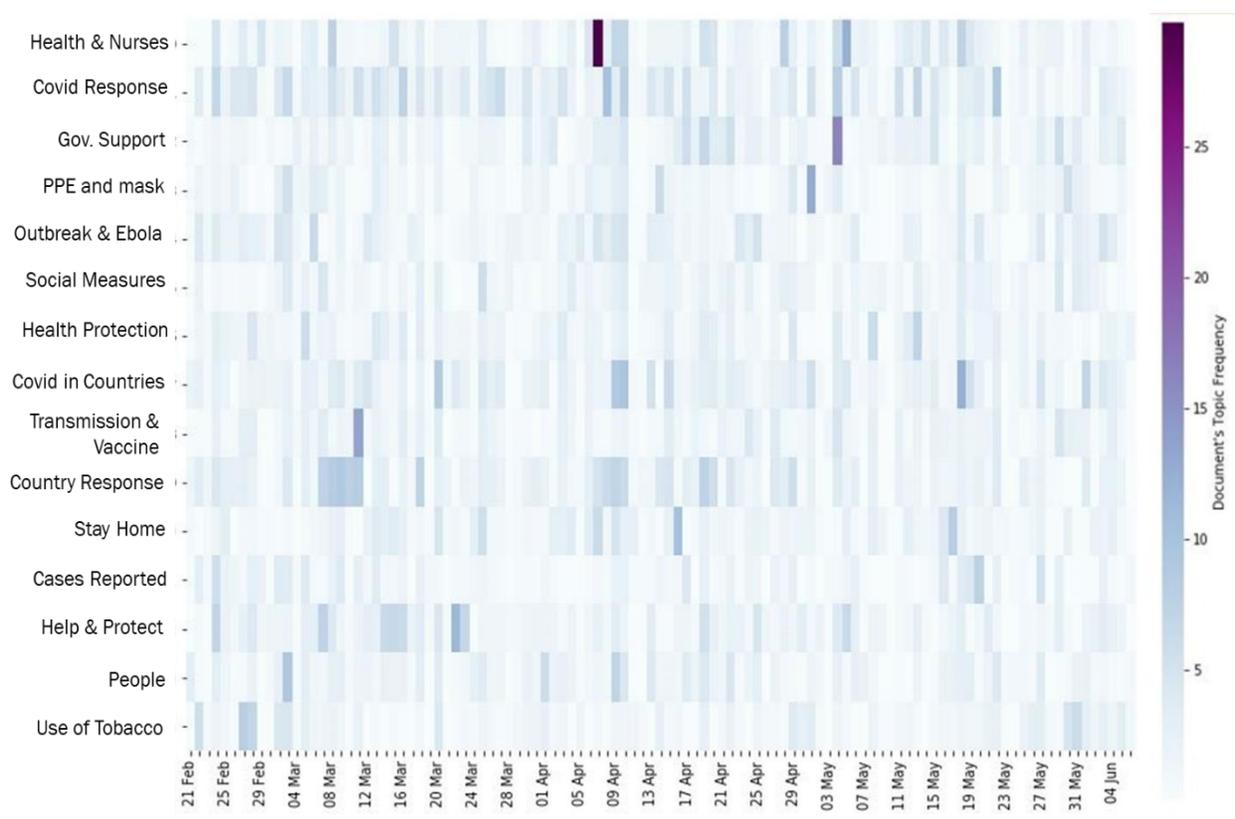

Dynamic Topic Model- WHO

**Figure 1a. Topic Frequency over time from WHO Tweets**

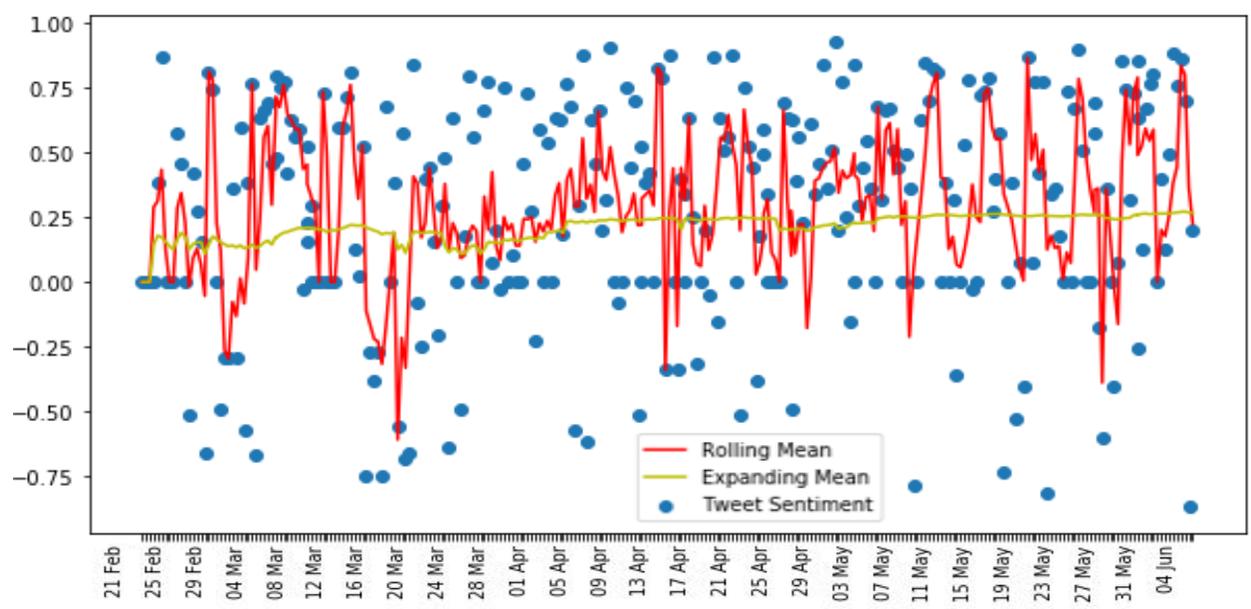

**Figure 1b. Sentiment Analysis over time from WHO Tweets**

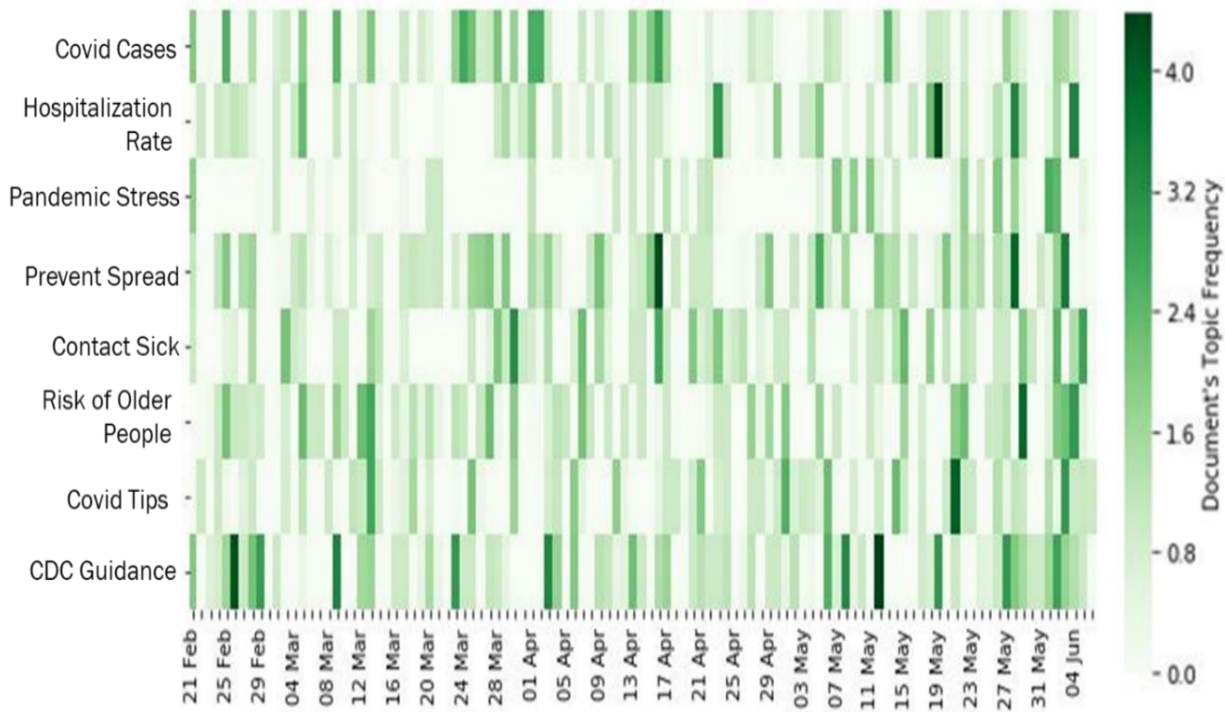

**Figure 2a. Topic Frequency over time from CDC Tweets**

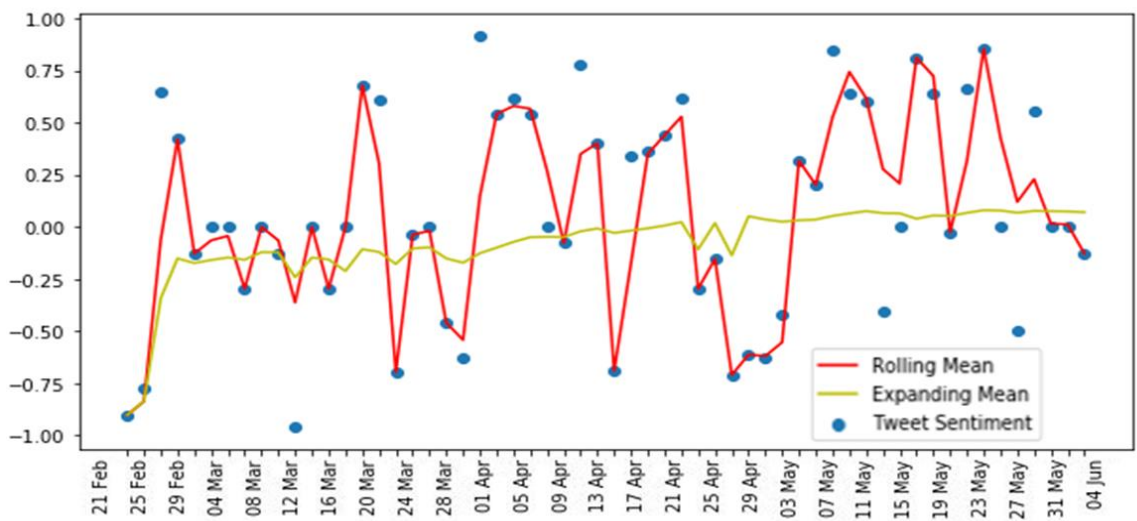

**Figure 2b. Sentiment Analysis over time from CDC Tweets**

**FEMA Interactions**

From around 2000 tweets of FEMA, 10 optimum number of topics frequency have visualized in Figure 3a. FEMA discussed about the importance of critical information in April, need of COVID-19 responses in mid-March and also after one month (mid-April), spreading of the virus at the end of April and need of supplies and food in April and May. Hurricane season also got attention of FEMA during the pandemic. Besides, FEMA emphasized on the importance of COVID-19 testing from April to May which is not mentioned by WHO and CDC. Also, lack of medical supply got notified from end of the March, April and May which shows after each month medical supply in US is needed.

By interpreting Figure 3a and 3b together, the extreme negative sentiments are observed when the need of medical supplies, lack of food and the emergence of upcoming hurricane season along with the increased spreading of the virus are discussed. Then, positive sentiments have found while importance of COVID-19 testing, availability of responses, fund and spreading of critical information got more importance.

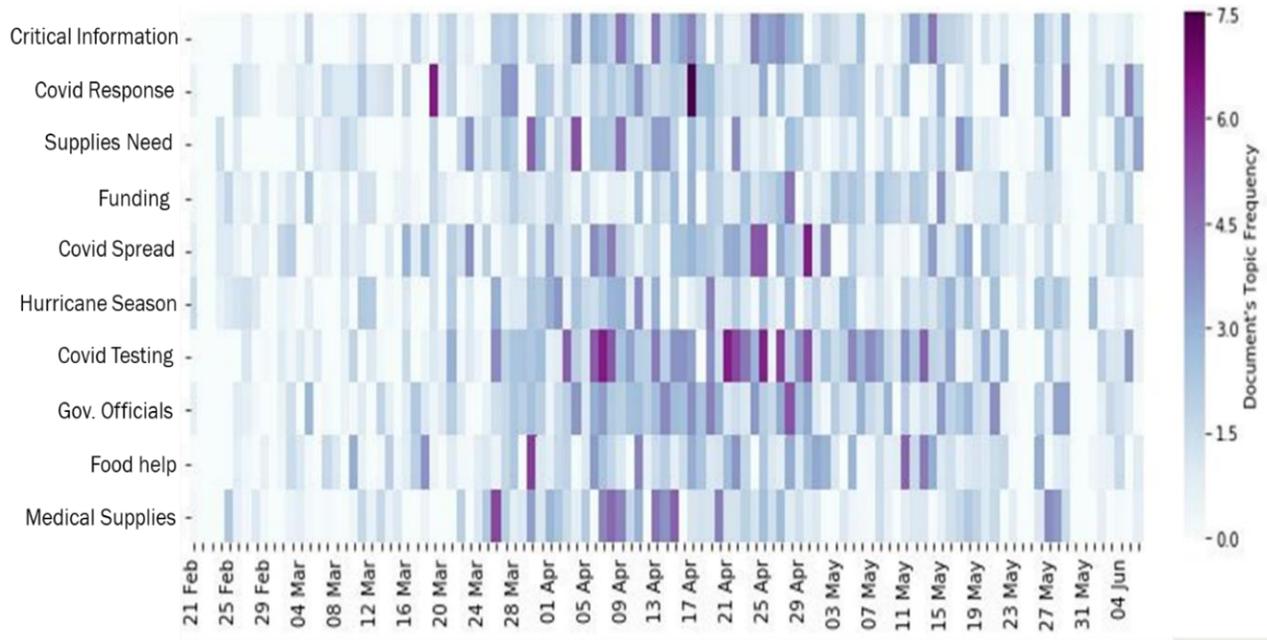

**Figure 3a. Topic Frequency over time from FEMA Tweets**

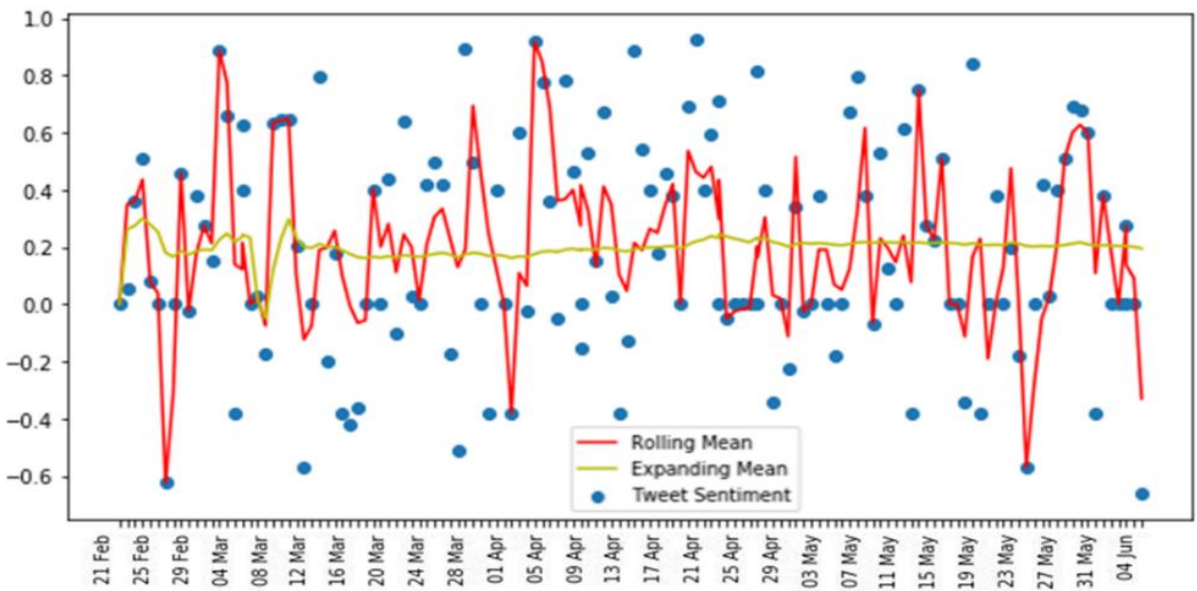

**Figure 3b. Sentiment Analysis over time from FEMA Tweets**

**FDOT Interactions**

FDOT tweets showed 8 optimum topics which are envisaged over time in Figure 4a. FDOT emphasized on working together to be successful over COVID-19 from May, travel and COVID-19 related information in March, use of airports in the beginning of April, roadway construction work consistently, importance of social distancing from April and the spreading of coronavirus in Floridians consistently.

By comparing Figure 4a and 4b, spikes in positive sentiments have found when the importance of social distancing, effectiveness of working together throughout the pandemic and restricted roadway construction work have discussed. In the other hand, extreme negative sentiments have observed while alarming spread of COVID-19, increase in number of cases among Floridians and use of airport topics became more frequent in Twitter.

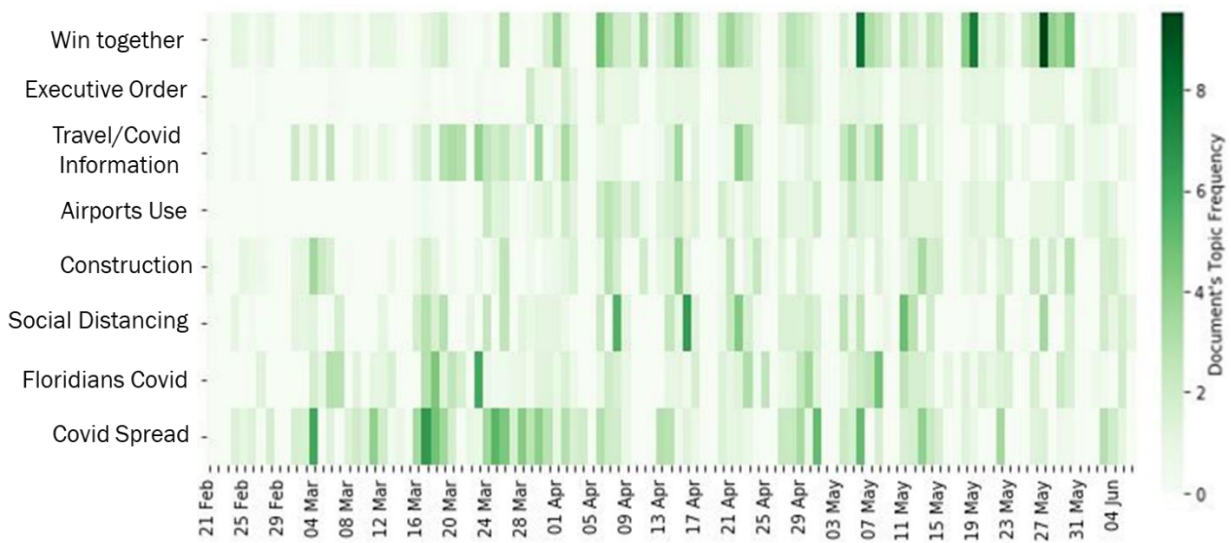

**Figure 4a. Topic Frequency over time from FDOT Tweets**

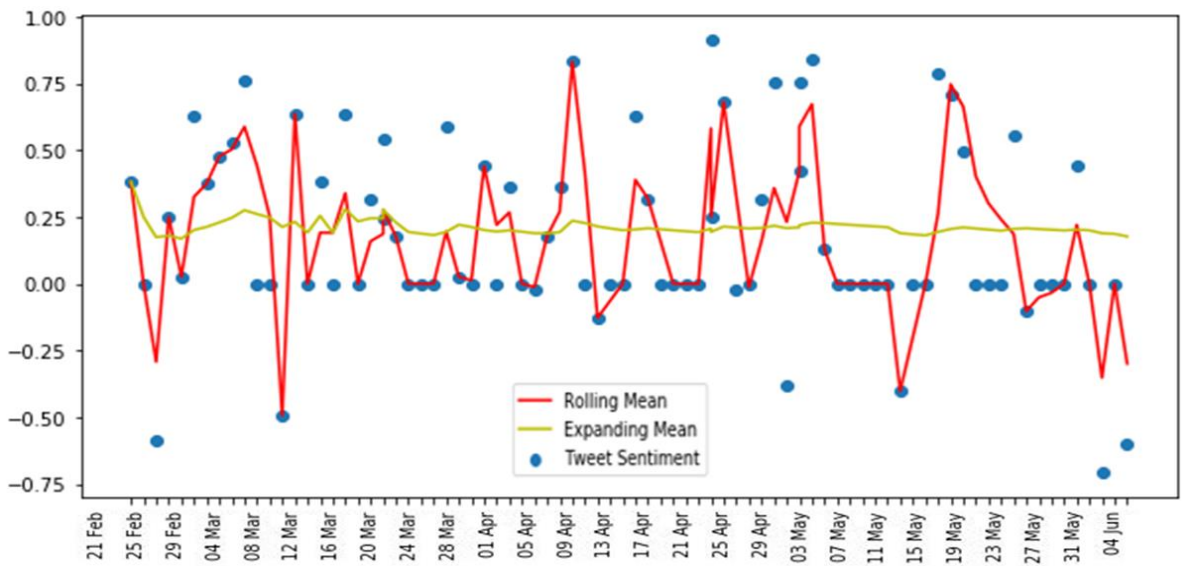

**Figure 4b. Sentiment Analysis over time from FDOT Tweets**

**DISCUSSION OF FINDINGS**

To identify the specific interaction patters along with the sentiments which caused the changes (increase and decrease) in the number of cases and deaths in US, the topic dynamics and sentiments over time are represented along with the number of cases and deaths of COVID-19 from Worldometer data (*24*) in Figure 5a and Figure 5b as following. From both Figure 5a and 5b, it is obvious that the number of cases and deaths started to increase rapidly from 20 March and hit the first pick (around 40,000 cases and 2,500 deaths per day) around 10 April. In this time duration, WHO discussed more about the lack of social measures, responses, requirement of adequate number of nurses, importance of wearing face masks; CDC emphasized on need of contact tracing, failure of preventing the spread and suggesting to follow their guidelines; FEMA mentioned about shortage of medical supplies, responses and need of critical information; and FDOT stated about the use of airports, continuous roadway construction works and lack of social distancing. All of these topics contribute to the increase in number of cases and deaths. If the entities put importance on these topics at the beginning of the pandemic (January 2020), then the number of cases and deaths could be much lower.

Then, the number of cases and deaths remained nearly constant from 11 April to 10 May. In this time, importance of social distancing, increase in funding, response and support from government, essence of pandemic stress management, new measures to prevent the spread and the requirement of PPE have started becoming into the discussion focus. The result of emphasizing on these issues is obvious from the last portion (11 May to 6 June) of the graphs (Figure 5a and 5b) where both the number of cases and deaths decreased. In this timeline, increased response from people and government, awareness about the lowering hospitalization rates, following health guidance, understanding the risk of elderly people, increased medical supply, and working together also helped to reduce the number of cases and deaths which results in flattening the curve.

The text-based infographics showed in previous sections have revealed different interaction patterns along with the sentiments from the four major public agencies during the pandemic. These interactions were actually generated from most frequently appeared words in different topics form these entities which are listed below in different timeline. The entire time is divided into four sub-period based on the lockdown (13 March) and reopening (beginning of May) time in US-

After analyzing all the results and infographics, the noteworthy findings are listed in below-

- *All the organizations discussed about the increased number of COVID-19 cases, need of responses from both people and government and spreading of the virus.*
- *The unique topics WHO discussed are the importance of PPE and wearing masks, home quarantine, need of nurses and vaccine, emergence of community transmission and the both negative and positive use of tobacco.*
- *CDC specifically mentioned about the importance of contact tracing, pandemic stress management and the negative impact of increased hospitalization rates.*
- *FEMA is the only organization who emphasized on the importance of COVID-19 testing, lack of medical supplies and the emergence of upcoming hurricane season along with the pandemic.*
- *FDOT put importance on social distancing, negative impact of using airports and roadway construction works.*

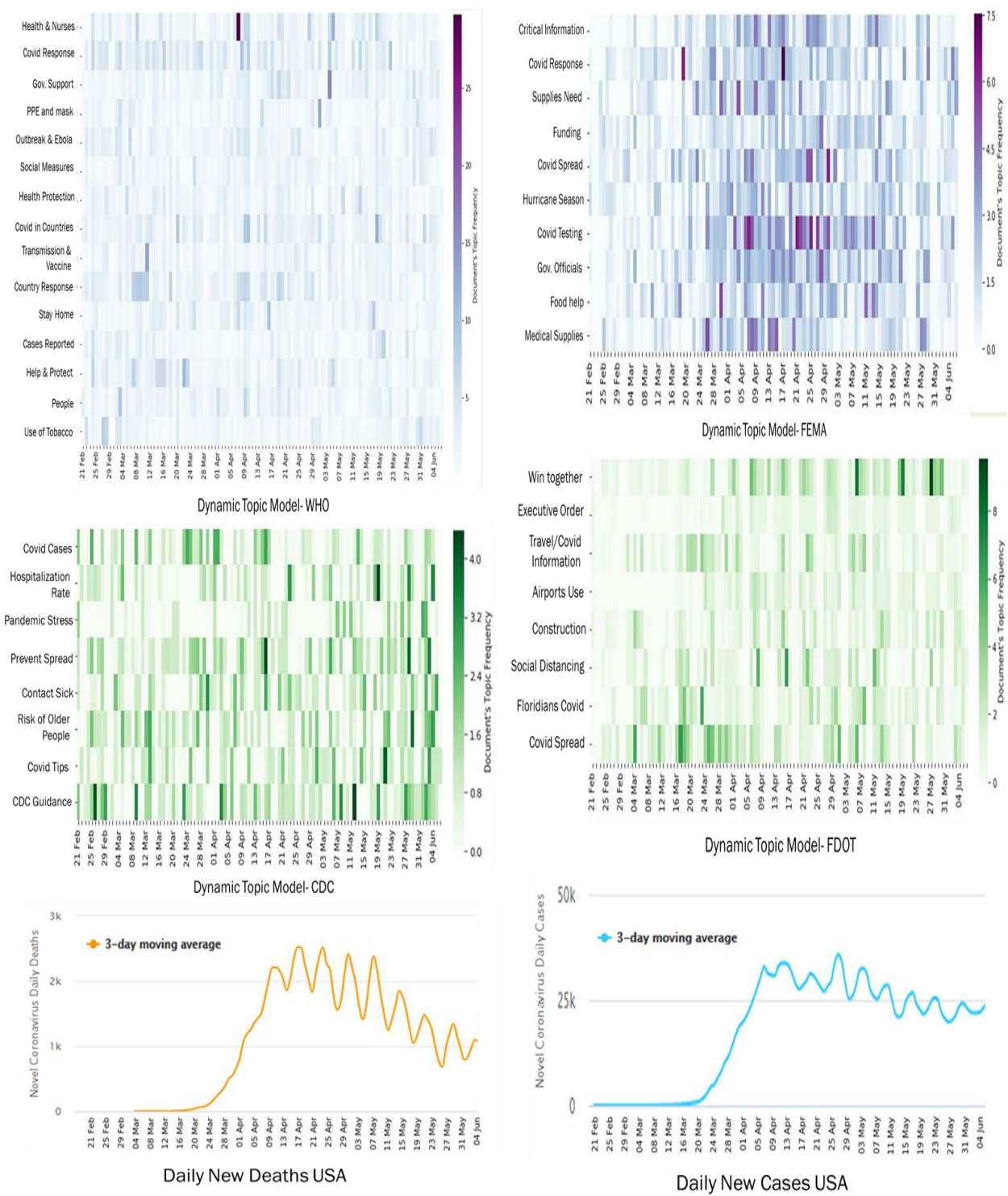

**Figure 5a. Organizations' Topic Dynamics with daily new Cases and Deaths in USA**

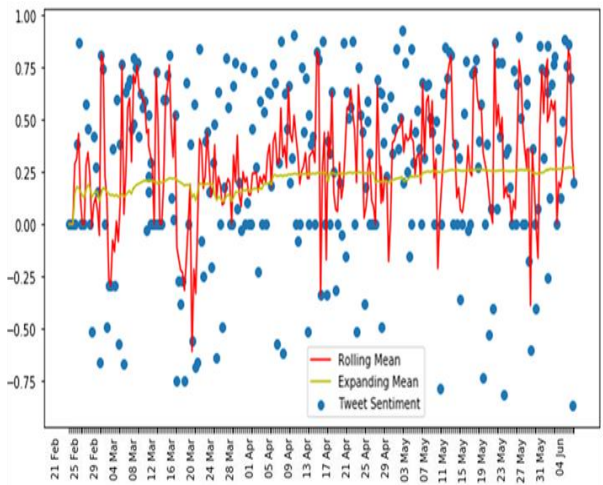

Sentiment Over Time- WHO

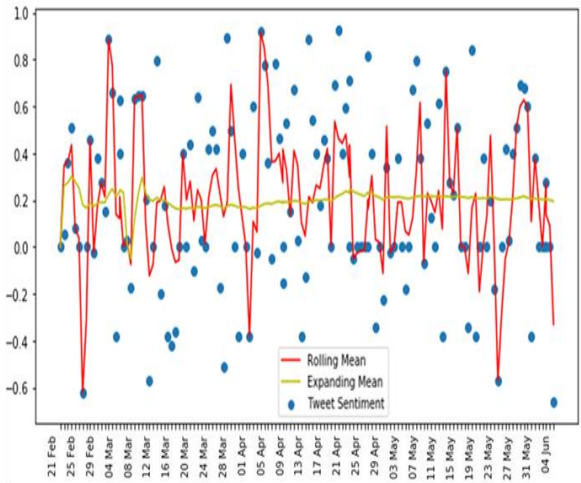

Sentiment Over Time- FEMA

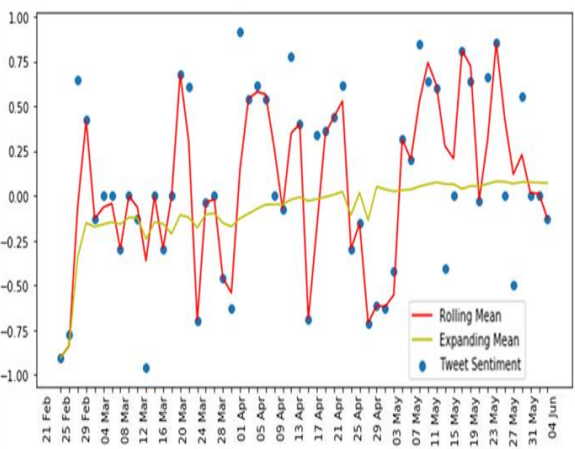

Sentiment Over Time- CDC

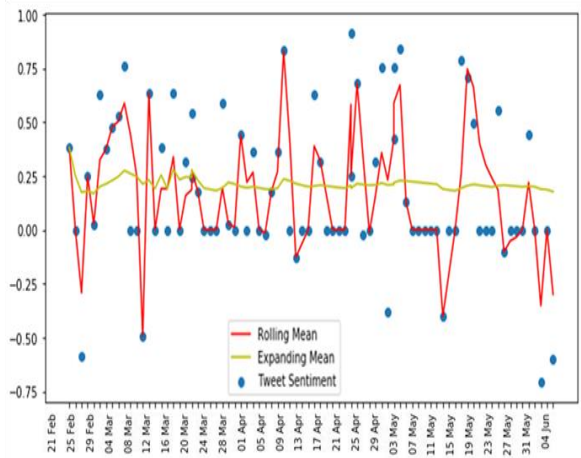

Sentiment Over Time- FDOT

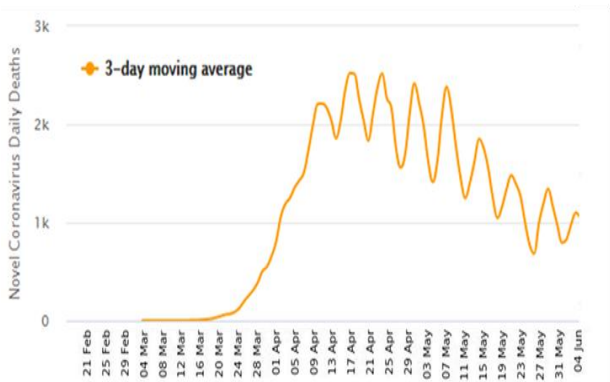

Daily New Deaths USA

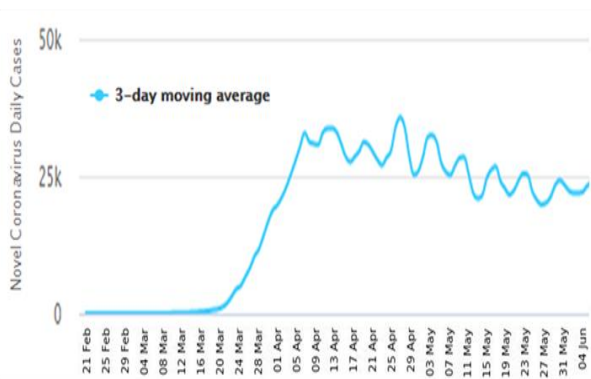

Daily New Cases USA

**Figure 5b. Organizations' Sentiment with daily new Cases and Deaths in USA**

**Table 1. Most Common Words from Organizations in specific timelines**

| Organizations | 21 Feb-19 March | 20 March- 10 April | 11 April-10 May | 11 May-06 June |
|---|---|---|---|---|
| | Before the lockdown | Beginning of the lockdown | During the lockdown | Re-opening phase |
| **WHO** | global, media, support, minister, personal, covid, increase, protect, health, safe, measures, countries, solidarity, community, transmission, cases, stop, now, spread, safehands, response, prevent | coronavirus, protective, medical, equipment, nurses, midwives, violence, economic, social, support, stay, home, confirmed, masks | world, international, response, fight, prime, emergency, advice, committee, tools, europe, physical, active, information, virus, infection | health, outbreak, ebola, human, public, tobacco, use, healthyathome, vaccines, solidarity, pandemic |
| **CDC** | health, US cases, doctor, people, risk, hands, coronavirus, CDC, help | facebook, coca, spread, older, travel, home, learn, CDC | covid, age, protect, home, help, gas, spread, states, report | cases, hospitalization rates, covidview, pandemic, stress, slow, guidance, learn, help, protect |
| **FEMA** | information, critical, need, emergency, firefighters, support, disaster, emotional, risk, emergency management, prevent, state, local, offcials | supplies, need, equipment, social, spread, medical, help, federal, guard, hospital, states, national | people, working, right, team, care, response, slodiers, response, medical, supplies, covid, million, funding | local, state, million, grants, facts, distancing, hurricane, season, tropical, storm, testing, health, food, national, members |
| **FDOT** | project, order, florida, road, covid, public, staff | social, distancing, executive, order, issued, travel, follow, visit, emergency, service, home, floridians | information, win, together, survey, stay, construction, work, traffic, driving, florida, essential, safe, practice | checkpoints, coordination, traveler, airports, enforcement, app, download, state, transportation, searching |

- *Only WHO discussed about the importance of wearing face mask on April and FEMA emphasized on COVID-19 testing on May which should be focused earlier and contributed to the initial increase of the number of cases and deaths in US.*
- *Before the lockdown time (Table 1), organizations discussed about community transmission, spreading of the virus, emergency management and health concerns along with the need of doctors.*
- *Just after the lockdown (20 March to 10 April), organizations emphasized on shortage of nurses, medical supplies, social distancing and contact tracing, which contributed to the Sudden jump in cases and deaths.*

- *During the lockdown (11 April-10 May), public agencies highlighted about the importance of social distancing, pandemic stress management, PPE requirement and the need of Government support to manage the pandemic.*
- *In the reopening phase (11 May to 06 June), organizations focused on managing hospitalization rate, risk of older people, adequate medical supplies and following the helath guidance, which eventually helped to reduce the number of cases and deaths.*

This study identified specific topics in Twitter which influenced the change in number of cases and deaths across US. Findings could help policy makers to take better preparation for the next steps of the ongoing pandemic as well as in future pandemic situations.

**CONCLUSIONS ANF FUTURE RECOMMENDATIONS**

Social media is considered as an effective information dissemination platform and showed prevalence in recent times. The influence of information received through social media on public behavior and decision-making is remarkable. However, spread of misinformation and information overload may impede public risk perception and personal protective actions. Exploring the origins of such information in times pandemic are inevitable, as such, Twitter interactions from agencies such as WHO, CDC, FEMA and FDOT were observed in this study in the emergence and outbreak of the novel coronavirus (COVID-19). A total of 8,600 tweets were analyzed from these four major public organizations. Several machine learning techniques such as (i.e. topic mining and sentiment ratings over time) are applied here to identify the dynamics of emergent topics during this unprecedented time. WHO was found to be the most active in social media during the first three and half month of COVID-19 pandemic as compared to the other agencies.

Results also indicate that most agencies were less vocal on the importance of face covering except for WHO in April which could have brought into attention earlier. Lack of healthcare professional, medical supplies, contact tracing, and social distancing may have contributed to the sudden increase in both number of cases and deaths in US. However, the importance of virus testing was emphasized more by FEMA along with the concerns about the upcoming hurricane season. CDC focused more on pandemic stress management and monitoring the hospitalization rates. FDOT discussed more about COVID situation in Florida, travel information, use of airports and construction during the pandemic. Finally, higher levels of social media interactions from agencies on social distancing and vulnerability of older people seemed to have contributed in reducing number of cases and deaths in US.

The findings of this research can support public health, emergency management, transportation and other agencies more efficient transfer of risk and response information as communities shift to new normal as well as in future pandemics. This can be done by identifying more effective information dissemination strategies for diverse user groups based on their social network characteristics, activities, and interactions in response to similar public health hazards. This study identified specific Twitter interaction topics from major agencies which may have influenced the community response and disease outbreak indicators of the pandemic. The methodologies, and implications of this research can be transferred in designing effective intervention policies to other natural and man-made disaster contexts in which public health risks become major concerns. The study considered up to 3,200 tweets from the four agencies due to the limitation set by Twitter on historical tweets, future studies should consider more agencies and more recent timeline.

## ACKNOWLEDGMENTS

The authors are grateful to National Science Foundation for the Rapid Response Research grant IIS-2027360 to support the research presented in this paper. However, the authors are solely responsible for the findings presented in this study.

## AUTHOR CONTRIBUTIONS

The authors confirm the contributions to the paper as follows: study conception and design: M. A. Ahmed, A. M. Sadri; data collection: M. A. Ahmed, A. M. Sadri; analysis and interpretation of results: M. A. Ahmed, A. M. Sadri; draft manuscript preparation: M. A. Ahmed, A. M. Sadri, M.H. Amini. All authors reviewed the results and approved the final version of the manuscript.